\documentclass[final,authoryear,5p,times,twocolumn]{elsarticle}

\usepackage{amssymb}

\journal{Icarus}
\bibpunct{(}{)}{;}{a}{}{,}
\usepackage{graphicx}

\newcommand{\arcsec}{\mbox{\ensuremath{^{\prime\prime}}}}

\usepackage{ulem}
\usepackage{pstricks}

\newcommand{\rem}[1]{}

\begin{document}

\begin{frontmatter}

\title{A spectral comparison of (379) Huenna and its satellite}

\author[mit,obs]{Francesca E. DeMeo\corref{irtf}}
\author[esac,obs,par]{Beno\^{i}t Carry}
\author[fm]{Franck Marchis}
\author[imc]{Mirel Birlan\corref{irtf}}
\author[mit,imc]{Richard P. Binzel}
\author[sjb]{Schelte J. Bus}
\author[imc]{Pascal Descamps}
\author[imc,an]{Alin Nedelcu\corref{irtf}}
\author[cal]{Michael Busch} 
\author[esac]{Herv\'e Bouy}

\address[mit]{Department of Earth, Atmospheric, and Planetary Sciences, Massachusetts Institute of Technology, 77 Massachusetts Avenue, Cambridge, MA 02139 USA}
\address[obs]{LESIA, Observatoire de Paris, 5 Place Jules Janssen, 92195 Meudon Principal Cedex, France}
\address[esac]{European Space Astronomy Centre, ESA, P.O. Box 78, 28691 
Villanueva de la Ca\~{n}ada, Madrid, Spain}
\address[par]{Universit\'e Paris 7 Denis-Diderot, 5 rue Thomas Mann, 75205 Paris CEDEX, France}
\address[fm]{University of California at Berkeley, Dept. Of Astronomy, 601 Campbell Hall, Berkeley, CA 94720 USA}
\address[imc]{IMCCE, UMR8028 CNRS, Observatoire  de Paris, 77 avenue Denfert-Rochereau, 75014 Paris Cedex, France}
\address[sjb]{Institute for Astronomy, 640 N. AÕohoku Place, Hilo, HI 96720 USA}
\address[an]{Astronomical Institute of the Romanian Academy, 5 Cu\c titul de Argint, RO-75212 Bucharest, Romania}
\address[cal]{Department of Earth and Space Sciences, University of California Los Angeles, 595 Charles Young Dr. E., Los Angeles, CA 90095 USA}

\cortext[irtf]{Visiting Astronomer at the Infrared Telescope Facility, which is
operated by the University of Hawaii under Cooperative Agreement
no. NNX-08AE38A with the National Aeronautics and Space
Administration, Science Mission Directorate, Planetary Astronomy
Program.}

\begin{abstract}
  We present near-infrared spectral measurements of Themis family
  asteroid (379) Huenna (D$\sim$98 km) and its 6 km satellite using SpeX
  on the NASA IRTF. The companion was farther than 1.5\arcsec~from the
  primary at the time of observations and was approximately 5 magnitudes
  dimmer.
  We
  describe a method for separating and extracting the signal of a
  companion asteroid when the signal is not entirely resolved from the
  primary. The spectrum of (379) Huenna has a broad, shallow
    feature near 1 $\mu$m and a low slope, characteristic of C-type
    asteroids. The secondary's spectrum is consistent with the taxonomic classification of C-complex or X-complex. The quality of the data was not sufficient to  identify any subtle feature in the secondary's spectrum.

\end{abstract}


\end{frontmatter}

\section{Introduction}
\small

Since the discovery of the small moon Dactyl around (243) Ida by the
Galileo spacecraft more than a decade ago, many more systems have been
discovered by either indirect methods  \citep[lightcuves,
  \textsl{e.g.}][]{Pravec2006} or direct detection \citep[direct
  imaging or radar echoes, \textsl{e.g.}][]{Merline1999, Margot2002,
  Ostro2006, Marchis2008}. The study of the physical and orbital
characteristics of these systems is of high importance, as it provides
density measurements and clues on formation processes through models
\citep[\textsl{e.g.}][]{Durda2004,Walsh2008}.  Primary formation
scenarios include capture (from collision or close approach), and
fission and mass-loss from either YORP spin up for small binary
asteroids \citep[D$<$20 km,][]{Pravec2007,Walsh2008} or from oblique
impacts for large asteroids \citep[D$>$100 km,][]{Descamps2008}.

Spectroscopic measurements of each component of multiple systems are
fundamental to understand the relationship between small bodies and their
companions. From high quality measurements, differences in slope and
band depths provide information on composition, grain sizes, or
effects of space weathering on each surface and help constrain
formation scenarios. Obtaining photometric colors or spectra of the
secondaries, however, is rarely possible due to the small (sub
arcsecond, often less than a few tenths of an arcsecond) separation
and relative faintness (often several magnitudes dimmer) of the
satellites. Photometric colors of TNO binaries, which are typically of
comparable size, have revealed their colors to be the same and are
indistinguishable from the non-binary population \citep{Benecchi2009}.
The first spectroscopic measurements of a resolved asteroid system,
(22) Kalliope, are presented by \citet{Laver2009} using the Keck AO system and its integral field spectrograph OSIRIS. They find the
spectra of the primary and secondary to be similar, suggesting a
common origin. More recently \citet{Marchis2009} presented a comparative spectroscopic analysis of (90) Antiope based on data collected using the 8m-VLT and its integral field spectrograph SINFONI. This binary asteroid is composed of two large (D$\sim$85 km) components. A comparison of Antiope's component spectra suggest that the components are identical in surface composition and most likely formed from the same material.

In November 2009, main-belt asteroid (379) Huenna reached perihelion,
and the system had a separation of over 1.5\arcsec. The primary is a
large \citep[D$\sim$98 km,][]{Marchis2008,Tedesco2002}
 C-type \citep{Bus2002b} and a member of the Themis family
\citep{Zappala1995}. Its secondary (S/2003 (379) 1) was discovered by
\citet{Margot2003}. With an estimated diameter of 6 km, the secondary is $\sim$5 magnitudes fainter at optical
wavelengths than the primary, and it has a highly eccentric orbit (e
$\sim$ 0.22) with a period of 87.60 $\pm$0.026 days
\citep{Marchis2008}. This irregular orbit suggests mutual capture from
disruption of a parent body or even an interloper \citep{Marchis2008}. It is not likely that the secondary formed by fragmentation of Huenna due to the YORP spinup effect, because the primary is too large for this process to be efficient \citep{Rubincam2000,Walsh2008}.
Here we present near-infrared spectral measurements of the primary and
secondary and as well as a new spectral extraction method to separate them.

\section{Observation and Data Reduction}

\subsection{Spectroscopy}

\begin{figure}
\centering
\includegraphics[width=0.43\textwidth]{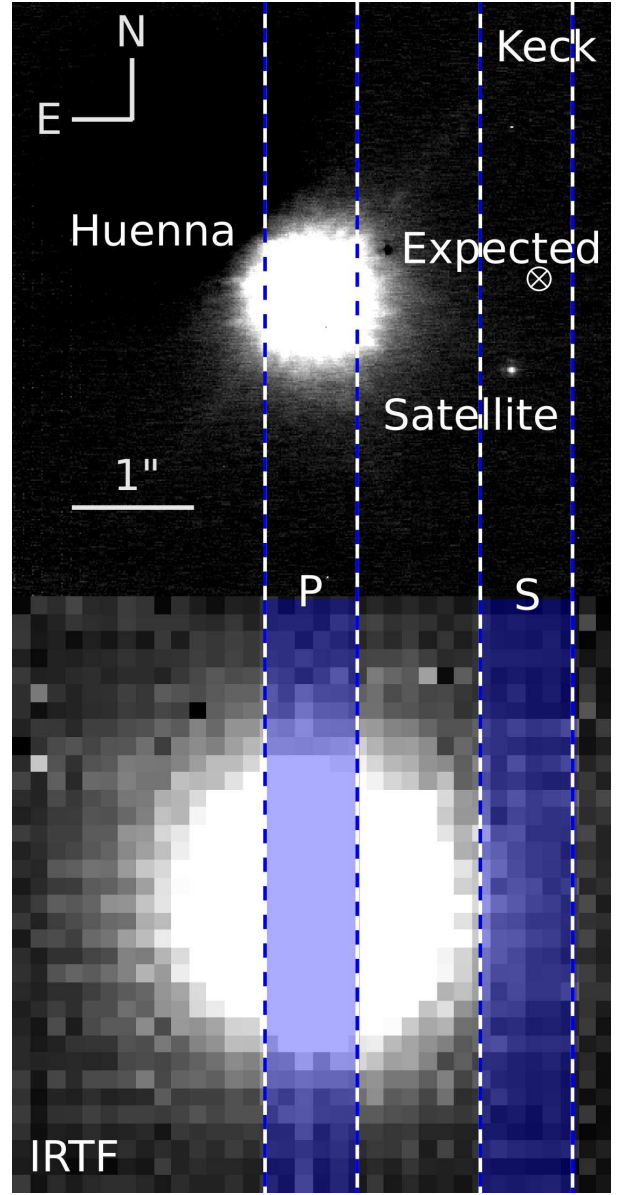}
  \caption[Slit position]{%
  Observing configuration:
  P marks the position of the slit for observations of the primary using the IRTF observations. S
  marks the position for the secondary which was chosen based on the expected position of the secondary.
  \textsl{Top}:  Adaptive-optics image obtained with NIRC2 at the
  Keck telescope.
  \textsl{Bottom}: Image obtained at the IRTF with SpeX slit-viewer, displayed at the same scale.
  Both images are oriented with north up and east left. 
  The plate scale of the IRTF spectrograph and IRTF guider
  (slit-viewer) are 0.15 and 0.12\arcsec/pixel, respectively.
} 
\label{slitview}
\end{figure}

  We observed the (379) Huenna system on 2009 November 28 UT, and
  obtained near-infrared spectral measurements from 0.8 to 2.5 $\mu$m
  using SpeX, the low- to medium-resolution near-IR spectrograph and
  imager \citep{Rayner2003}, on the 3-meter NASA IRTF located on Mauna
  Kea, Hawaii.  Observations were performed remotely from the Centre
  d'Observation \`a Distance en  
  Astronomie \`a Meudon \citep[CODAM,][]{Birlan2004} in Meudon,
  France. Using the Binary Orbit Fit (BOF) algorithm from
  \citet{Descamps2005} and the orbital parameters provided in
  \citet{Marchis2008}, the secondary was expected to be located about
  1.88\arcsec~W and 0.17\arcsec~N with respect to the primary at the
  time of our observations.
The configuration of the slit (shown in Fig.~\ref{slitview}) was
  thus chosen as following. The
  slit was first centered on the primary (position ``P'' in
  Fig.~\ref{slitview}) and 8 spectra of 120 seconds were taken,
  alternated between two different positions (usually noted as `A' and
  `B') on a 0.8\arcsec$\times$15\arcsec~slit aligned in the
  north--south direction. We chose not to align the slit along
    the parallactic angle to facilitate the positioning of the slit on
  the secondary for these challenging observations. Next, the slit
  was shifted west $\sim$1.8\arcsec~(15 pixels; position ``S'' in
  Fig.~\ref{slitview}) to maximize the flux from the bulge on the
  southwest portion of the guider image, the secondary, and minimize
  contamination from the light of the primary. We continued to guide on
  the primary while taking spectra of the secondary. Because of the long
  period ($\sim$87 day) of the secondary, its motion relative to the
  primary during our observations was negligible. We spent two hours
  taking spectra of 120 seconds of the secondary, although we combine
  and present only the 11 spectra toward the end of the night, when the
  seeing was lowest and the signal of the secondary was detectable. The
  solar-type standard star Landolt 102-1081 \citep{Landolt1983} was
  observed to remove the solar contribution from the reflected light,
  and an argon arc lamp spectrum was taken for wavelength
    calibration.


The dome was closed for the first two hours of the observations due to
high humidity. During the beginning of the observations the seeing was
around 1\arcsec~and humidity was high. The data we present here,
however, were taken toward the end of the night when the humidity had
dropped to 10\% and the seeing was down to about 0.5\arcsec. 

Reduction was performed using a combination of routines within the
Image Reduction and Analysis Facility (IRAF), provided by the National
Optical Astronomy Observatories (NOAO) \citep{Tody1993}, and
Interactive Data Language (IDL). A software tool called
\texttt{Autospex} was used to streamline reduction procedures
to create a bad pixel map from flat field images, apply
a flat field correction to all images and perform the sky subtraction
between AB image pairs. The primary spectra were combined and the final spectrum was extracted with a
large aperture centered on the peak of the spectrum using the \texttt{apall}
procedure in IRAF.  The spectral extraction procedure of the satellite
is described in Section 3. A wavelength calibration was
performed using an argon arc lamp spectrum. The spectra were then divided by
that of the standard star to determine the relative reflectance and normalized to unity at 1.215 $\mu$m.

\subsection{Imaging}

On the same night (see Table~\ref{tab-obs}),
we also observed Huenna using the Adaptive Optics 
system of the Keck-II telescope and its NIRC2 infrared camera 
(1024$\times$1024 pixels with a pixel scale of 9.94 $\times$ 10$^{-3}$ 
arcsec/pixel). Two Kp-band (bandwidth of 0.351 $\mu$m centered on 2.124 
$\mu$m) images made of 4 coadded frames each with an integration time of 40 seconds 
were basic-processed using classical methods (sky subtraction, 
flat-field correction and bad pixel removal). In the resulting frame 
shown in Fig.~\ref{slitview} (averaged observing time 13:49:05), 
obtained after shift-adding these two frames, the primary is not resolved 
and slightly saturated. The satellite is detected with a peak 
SNR$\sim$23 and no motion with respect to the primary was detected over 
the 6 minutes of observation.

The position of the satellite as measured on the NIRC2 image
  (Fig.~\ref{slitview})
  is 1.67\arcsec$\pm$0.05\arcsec~W and
  0.62\arcsec$\pm$0.05\arcsec~S, 
  significantly offset from the predicted position based on the \citet{Marchis2008} orbital
  model (1.88\arcsec~W and 0.17\arcsec~N).
  This highlights the need for a refined period and adjustment for
  precession effects due to the irregular shape of the
  primary. Additional astrometric measurements are necessary to better
  constrain the orbital period and precession. For the spectroscopic observations on the IRTF, because the slit was oriented N-S and the displacement of the satellite in the E-W direction with respect to its expected position was only 0.2\arcsec~(compared to a 0.8\arcsec~slit), the slit was well-positioned on the satellite (see Fig.~\ref{slitview}).

\begin{table}[t]
\begin{center}
\caption[Observing log]{Observations on 28 November 2009}
\label{tab-obs}
\begin{tabular}{cccl}
\hline
\hline
Target & Time (UT) & Airmass & Instrument \\
\hline
(379) Huenna & 12:56-13:08 & 1.31-1.37 & Spex, IRTF\\
S/2003 (379) 1& 13:13-13:56 & 1.40-1.75& Spex, IRTF\\
L102-1081 & 14:19-14:26 &1.34-1.31& Spex, IRTF\\
(379) Huenna & 13:46-13:52 & 1.6-1.75& NIRC2, Keck-II\\
\hline
\end{tabular}
\end{center}
\end{table}

\section{Spectral Extraction of the Satellite}

\begin{figure}
\begin{center}
  \includegraphics[width=0.48\textwidth]{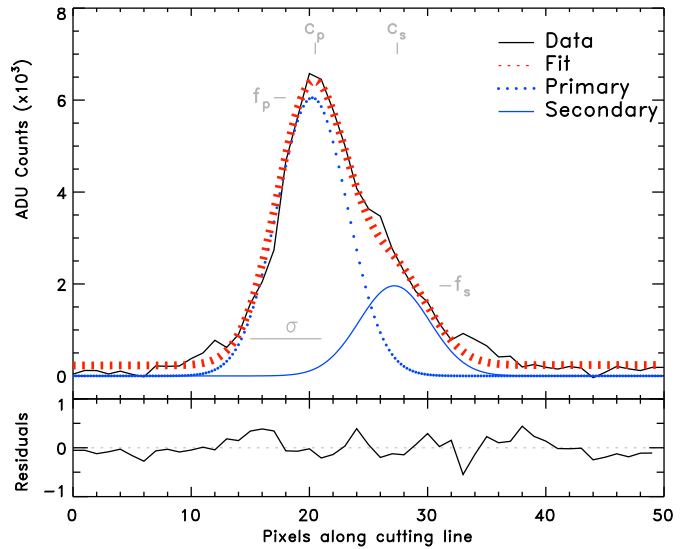}%
  \caption[Spectral extraction]{   A cut of the (379) Huenna spectrum along a wavelength-constant
    line ($\lambda = 1.23$~$\mu$m).
    Acquired data (observing time 13:58 UT) is plotted as a black
    solid line, and the corresponding best-fit curve is drawn as a
    dotted red line. Individual Gaussian functions for Huenna
    (primary) and its satellite (secondary) are drawn in bulleted
    and solid blue curves, respectively. Central positions ($c_p$, $c_s$),
    peak amplitudes ($f_p$, $f_s$), and 
    Gaussian standard deviation ($\sigma$),
    accordingly to notation in Eq.\ref{eq: fit},
    are labeled for clarity. Residuals of the fit are plotted on the bottom panel. }
  \label{fig: extract}
\end{center}
\end{figure}

 \indent Given the spatial resolution of SpeX, which is limited by the
  atmospheric seeing, and the angular separation of (379) Huenna and its
  satellite at the time of our observations ($\sim$1.5\arcsec),  the spectrum of the secondary was contaminated by diffuse light from the primary. We thus extracted both spectra (the companion and the diffuse light of the primary) at once, by adjusting a fit  function at  each wavelength composed of the sum of two Gaussian
  functions (one for each component) on top of a linear background: 

\begin{equation}
  \textrm{fit} =
        \frac{f_p}{\sqrt{ 2 \pi \sigma^2}}~.~\textrm{exp} \left(\frac{-(x-c_p)^2}{2\sigma^2}\right)
      + \frac{f_s}{\sqrt{ 2 \pi \sigma^2}}~.~\textrm{exp} \left(\frac{-(x-c_s)^2}{2\sigma^2}\right)
      + ax + b
  \label{eq: fit}
\end{equation}

  \noindent where $x$ describes the pixels perpendicular to the
  wavelength  dispersion direction. The 7 free parameters are:
  $a$ and $b$, the slope and absolute level of the background,
  $f_p$ and $f_s$, the maximal flux of each component centered on
  $c_p$ and $c_s$, respectively,
  and $\sigma$ the Gaussian standard deviation.
  In Fig.~\ref{fig: extract}  we present an example of the adjustment
  showing both components and their sum plotted with the data. \\ 
%
%
%
%
%
%
%

\begin{figure}
\begin{center}
  \includegraphics[width=0.48\textwidth]{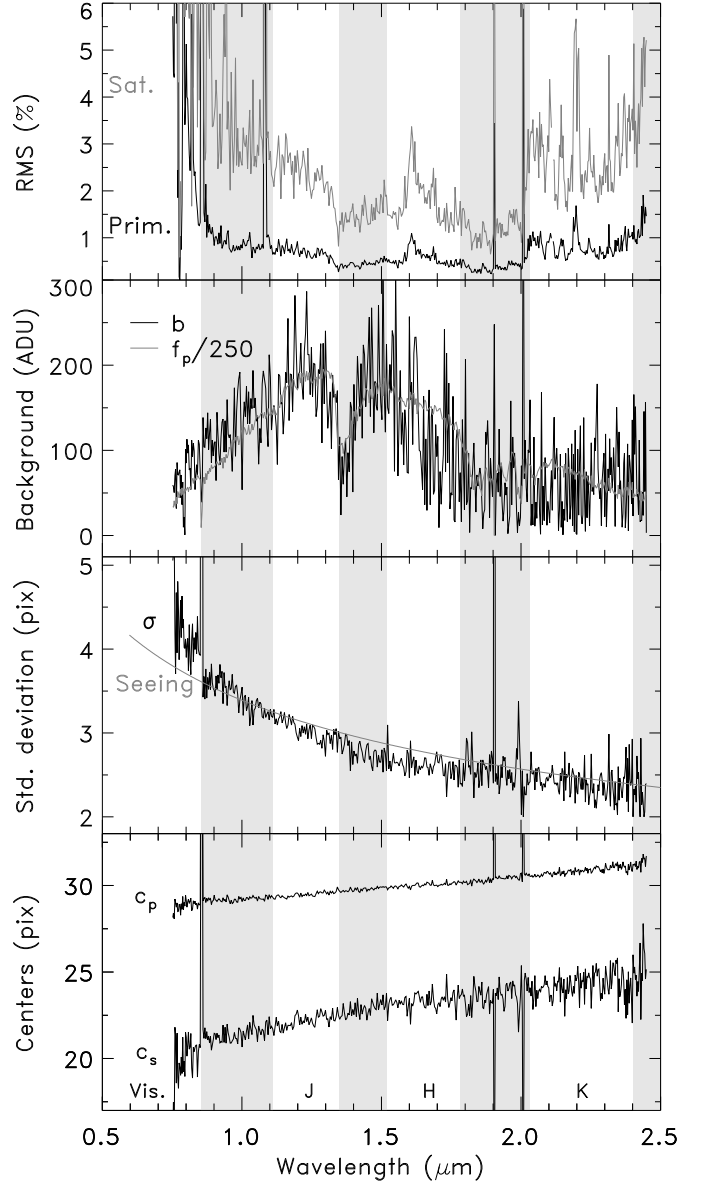}%
  \caption[Extraction Paramaters]{ Parameters of the spectral extraction plotted as function of the
    wavelength.The gray areas represent the wavelength ranges affected
    by telluric absorptions, and we labeled the atmospheric windows
    with their usual designation (visible, J, H, and K).
    \textsl{From bottom to top. Sub-panel 1:}
    position of the two Gaussian centers, Huenna ($c_p$) and its
    satellite ($c_s$).
    \textsl{Sub-panel 2:}
    the Gaussian standard deviation ($\sigma$)
    compared to the atmospheric
    theoretical seeing (see text).
    \textsl{Sub-panel 3:}
    the background level ($b$)
    compared to a fraction of Huenna flux ($f_p$).
    \textsl{Sub-panel 4:}
    the residual of the fit, displayed as percent of the peak value of
    the spectra ($f_p$ and $f_s$ for Huenna and its satellite respectively).
}
\label{fig: param}

\end{center}
\end{figure}

  \indent We verified the accuracy and robustness of the extraction by
  examining the residuals of the fit and the evolution of the free parameters according to
  wavelength shown in Fig.~\ref{fig: param}.
  First, the angular separation between the two components, the peak of the diffuse primary light and the secondary,
  ($c_p$$-$$c_s$) is nearly
  constant, with a value of $1.04 \pm 0.25$\arcsec, the uncertainty
  being mostly due to the uncertainty of the position of the secondary
  ($c_s$, see Fig.~\ref{fig: param}).
  This separation is the order of magnitude expected
  (0.62\arcsec~along the North-South direction, see
  Fig.~\ref{slitview}).\\
  \indent Second, the Gaussian standard deviation ($\sigma$) follows
  the expected trend due to the seeing power-law dependence with
  wavelength. 
  Indeed, the seeing is proportional to $\lambda^{-1/5}$, which means
  that the amount of diffused light from Huenna entering the slit is not
  the same at short (\textsl{e.g.} J) and long (\textsl{e.g.} K)
  wavelengths. The same effect occurs with the slit-dispersed light: in the direction 
  perpendicular to the wavelength dispersion, the spectrum is broader at shorter wavelengths
  and narrower at longer wavelengths. The combined effect is
  a $\lambda^{-2/5}$ dependence of the standard deviation
  ($\sigma$). We plot in Fig.~\ref{fig: param} the theoretical
  wavelength dependence function of the Gaussian standard deviation,
  labeled ``\textsl{seeing}'', for a seeing value of 0.5\arcsec~at 0.5 $\mu$m. \\
\indent We set the standard deviation of the satellite Gaussian function
  equal to Huenna's to simplify the fitting algorithm.  However,
  from the discussion above, we can expect a wavelength dependence in 
  $\lambda^{-1/5}$ for the satellite (which is entirely covered by
  SpeX slit).
  This implies we slightly underestimate the size of the satellite
  Gaussian function toward longer wavelength, and thus the satellite
  flux, and therefore its spectral slope. \\
  \indent Third, the peak-to-peak amplitude measured
  for Huenna and its
  satellite corresponds to the value of $\Delta m \approx 6$
  reported by \citet{Marchis2008}:
  the ratio $f_s/f_p$ measured for the spectra is
  $0.4 \pm 0.1$, and the amount of diffused light from Huenna
  entering the
  slit is about 1\% of the overall peak amplitude of Huenna ($f_H$), calculated using a
  Gaussian profile for Huenna due to $\sim$0.5--0.6\arcsec~seeing,
  conditions similar to what we had at the end of the night.  Note that $f_H$ is Huenna's overall maximum, \textit{not} the 
  local peak we measure for dispersed light that we call $f_p$.\\
  \indent Finally, the background level ($b$), mimics the spectral
  behavior of Huenna, whose diffused light dominated the light entering the slit, as
  expected.
  The amount of diffused light entering the slit
  1\arcsec~from $c_p$ (a separation corresponding to 2\,$\sigma$) is
  about 10$^{-5}$ $\times f_H$, corresponding to $\approx f_p$/250
  (under the same seeing assumptions). 

We also test our extraction method by comparing the high quality
  spectrum of the primary while centered on the primary extracted by both
  methods. In the IRAF reduction, the center peak is identified, an
  aperture width is chosen, and all the flux within that region is
  summed. In our method we fit Gaussian functions to the flux. An
  additional error is introduced in our extraction method because the
  fit functions can slightly over or underestimate the flux. The
  largest error is introduced before 1\,$\mu$m and past 2.3\,$\mu$m as is
  shown in the top panel of Fig.~\ref{fig: param}. The spectra are
  generally consistent, although the slope of the spectrum from the
  new extraction method is lower (0.70$\pm$0.09 vs. 1.04$\pm$0.05) and
  we do not see the 1 micron band as clearly because of the poorer fit
  of the function at shorter wavelengths. This suggests that the
  errors in the slope from the Gaussian extraction are greater than
  the formal errors listed. The spectra of the diffuse primary and the
  secondary have a low signal-to-noise ratio and thus high errors due
  to the low flux and the limited time when the seeing was
  sufficiently low.
  
%
%
%

\section{Results}

\begin{table*}
\begin{center}
\caption{Slope Measurements}
\label{tab-slopes}
\renewcommand{\footnoterule}{}
\begin{tabular}{cccc} \hline
Object$^1$
 & Slope 0.9-2.4 $\mu$m  & Slope 0.9-1.3 $\mu$m& Slope 1.4-2.4 $\mu$m \\
\textbf{} & [\%/(10$^3$ \AA)]  & [\%/(10$^3$ \AA)]&
[\%/(10$^3$ \AA)] \\
\hline
                       (379) IRAF  &    1.04$\pm$0.05  &    0.70$\pm$0.02  &    1.29$\pm$0.02 \\
                       (379) Gauss   &    0.70$\pm$0.09  &   -1.20$\pm$0.05  &    0.86$\pm$0.05\\
                       (379) Diffuse &    2.34$\pm$0.59  &    2.53$\pm$0.38  &    1.45$\pm$0.38\\
                        S/2003 (379) 1   &    2.51$\pm$1.17  &    3.41$\pm$0.77  &    1.20$\pm$0.77\\
\hline
\end{tabular}
\end{center}
\end{table*}

\begin{figure}
\begin{center}
  \includegraphics[width=0.5\textwidth]{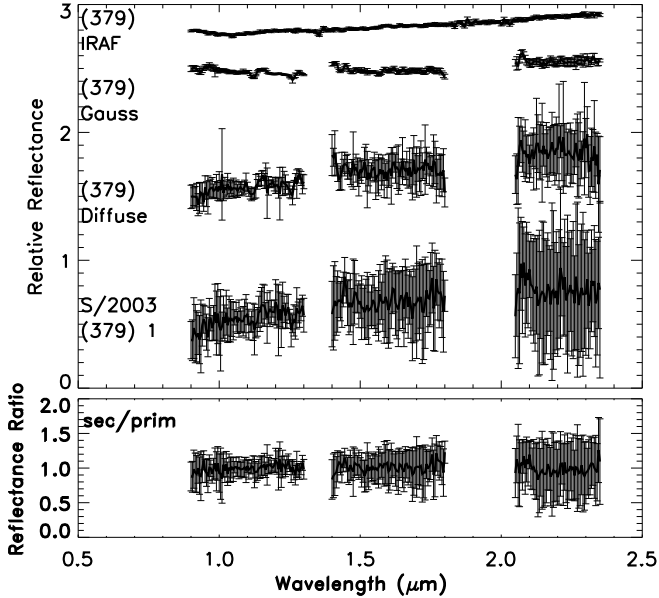}%
  \caption[Observed spectra]{Top section: From top to bottom: The spectrum of
      the primary with the slit centered on the primary reduced using
      IRAF; the same data reduced using the new Gaussian method presented here;
      the spectrum of the primary with the slit centered on the
      secondary measuring the diffused light; and the spectrum of the
      secondary. The spectra are plotted with one sigma errors and are
      normalized to unity at 1.215 $\mu$m and are shifted by
      -0.5,+1.5,+2.5, and +2.8, respectively. Bottom section: The
      ratio of the secondary and diffuse primary spectra both with the
      slit centered on the secondary.} 
  \label{fig:spectra}

\end{center}
\end{figure}

The spectra of Huenna and its companion are presented in
Figure~\ref{fig:spectra}.  Slopes are calculated between 0.9 and 2.4 $\mu$m using a chi-square
minimizing linear model (using the IDL routine \texttt{linfit}) and
are listed in Table~\ref{tab-slopes} with their formal errors for
various wavelength ranges. The telluric regions from 1.3 to 1.4\,$\mu$m
and from 1.8 to 2.0\,$\mu$m were excluded from the calculation. A
spectrum of the primary published by \citet{Clark2010} is taxonomically in
  agreement with our spectrum, also a C-type, however the band minima
  positions differ (1.020\,$\mu$m versus 1.045\,$\mu$m for our data). The
  slope of our spectrum is also lower than the previously published
  data, although they agree within the errors, 1.04$\pm$0.55 and
  1.94$\pm$0.53 \%/10$^3$ \AA. An additional error of 0.5 \%/10$^3$
  \AA~is added to these slopes because in this case we are comparing
  data from different nights using different standard stars.

The spectrum of the secondary is consistent with a C-complex or X-complex spectrum. The C- and X-complexes are broad umbrella terms for B-, C-, Cb-, Cg-, Cgh-, Ch-types and X-, Xc-, Xe-, and Xk- types, respectively. Without accompanying visible data, C-complex and X-complex spectra are indistinguishable except in the case of the presence of a broad 1 $\mu$m feature indicating C-type \citep{DeMeo2009}. As can be seen in Table~\ref{tab-slopes} the slopes of the spectra of the secondary and the diffuse light of the primary are greater than the slope of the primary spectrum even considering the large error. Particularly, the slopes are much greater in the shorter wavelength region, while the slopes are all relatively consistent in the higher wavelength region.  Because this slope effect is seen for both spectra, we suspect it is not due to any compositional differences but rather to the observational circumstances.

Differences in slope could be caused by atmospheric differential
refraction which varies as a function of wavelength and is stronger at shorter wavelengths
\citep{Filippenko1982}.  For our observations, the slit was oriented in the
  north-south direction rather than along the parallactic angle. We chose this orientation to facilitate the positioning of the slit on
  the secondary while reducing the contamination of diffuse light from the primary. However, this slit orientation could introduce slope effects. Airmass differences at the times of the observations are
  likely the biggest contributor to slope effects. The airmass was 1.3
  when centered on the primary and 1.4-1.7 when centered on the
  secondary.  Because the airmass of the standard star was 1.34 at the
  time of observation, effects of atmospheric refraction were not
  entirely removed for the observations when centered on the
  secondary. 
  
  Considering the effects of airmass and parallactic angle that introduce an additional slope affect to the data, we are unable to be certain whether the slope of the secondary is consistent with the primary.  This allows several possible formation scenarios. The satellite could be a fragment from a former parent body that was disrupted to form Huenna and the secondary. The satellite could be a member of the Themis family that was captured into an eccentric orbit around Huenna. Although more unlikely, it could also be a captured interloper of similar taxonomic type. We are confident, however, that the satellite is not of a significantly different taxonomic type. 
  
  We have presented the third resolved spectrum of a moonlet satellite
of a multiple asteroid system, the first being (22) Kalliope
from \citet{Laver2009} and the second, (90) Antiope
\citep{Marchis2009}. Our innovative extraction method allows us to
derive the spectra of the satellite (with a magnitude difference of 5
and distance of 1.5$\arcsec$) and the primary, and this method could
be used for similar future projects.  These results show the
capability of a medium-size telescope to aid in the
  measurement of multiple system asteroids with large
separations. Further observations with larger telescopes, particularly
those equipped with adaptive optics, can be used to perform the
similar measurements for multiple asteroid systems which are not
easily separable (angular distance $<$1$\arcsec$), and will be useful
to constrain formation scenarios.

\section{Acknowledgments}

We are grateful to Andy Boden, Gaspard Duchene, Shri Kulkarni, Bruce
Macintosh, and Christian Marois for helpful discussions and for
sharing results from their data. This research has made use of NASA's
Astrophysics Data System. FD y BC consideran que este proyecto fue una
\textsl{huenna} oportunidad para trabajar juntos. The authors
wish to recognize and acknowledge the very significant cultural role
and reverence that the summit of Mauna Kea has always had within the
indigenous Hawaiian community. We are most fortunate to have the
opportunity to conduct observations from this mountain.

\bibliographystyle{elsarticle-harv}

\end{document}